# Direct evidence of intercalation in a topological insulator turned superconductor

Tanmay Das, Somnath Bhattacharyya *, Bhanu Prakash Joshi, A. Thamizhavel and S. Ramakrishnan

Department of Condensed Matter Physics and Materials Science,

Tata Institute of Fundamental Research, Mumbai 400005, India

* Ph.: (+91) (22) 22782435; Fax: (+91) (22) 22804610/11;

*email: somnath@tifr.res.in

**ABSTRACT**

**A topological insulator $Bi_2Se_3$ which shows superconducting property due to Cu doping has been studied using high resolution transmission electron microscopy which provides for the first time, the direct evidence of intercalation sites. A careful study exhibits the presence of two intercalation sites in this material when projected along the [-1-1 2 0] direction. Stoichiometry of the intercalated unit cell was also calculated from the projected potential.**

**Keywords**

Topological Superconductor, intercalation, TEM, exit face wave reconstruction

Topological insulators are one of the most interesting candidates in contemporary research because of their remarkably different surfaces and bulk states. Due to strong spin-orbit coupling these materials show relatively longer range quantum entanglement which makes them suitable for building next generation quantum computational devices[1, 2, 3]. Amongst other such materials, $Bi_2Se_3$ is of particular importance due to its large bulk band gap and surface states with a single Dirac fermionic mode which shows a linear, lightlike dispersion[4]. The presence of the Dirac mode makes these surface states topologically secured from different dissipative interactions. These features promote it as a potential candidate for fault-tolerant quantum applications[5,6]. Recently $Bi_2Se_3$ came into the limelight by showing superconductivity when dopped with Cu[7]. While an intercalated $Cu_xBi_2Se_3$ shows a superconducting phase, Cu substituted structure $Cu_xBi_{2-x}Se_3$ does not exhibit any trace of superconductivity. This made researchers to believe that Cu intercalation into $Bi_2Se_3$ structure is crucial to make it superconducting[7]. It was also reported[7] that any short or medium range ordering in Cu intercalation was absent in the bulk material. Therefore it is possible to probe the intercalated structure only at the atomic scale. High resolution Transmission Electron Microscopy can be proved as a very useful tool to reveal the intercalated structure though no thorough investigation of this kind on high quality single crystals of $Cu_xBi_2Se_3$ is reported yet. Motivation of the present study is to bridge this gap. A high-quality single crystalline samples of superconducting Cu-intercalated $Bi_2Se_3$ was investigated here at atomic scale using Transmission electron microscopy (TEM). To extract true material specific information and to avoid other unwanted parameters introduced by electromagnatic lenses, phase of exit face wave is retrieved from through focal image series of superconducting $Cu_xBi_2Se_3$ to verify the presence of intercalation site/s in $Bi_2Se_3$.

Since single crystal of $Cu_xBi_2Se_3$ within $0.1<x<0.15$ was reported to exhibit superconductivity[7] so in present study using Bridgeman method a single crystal of $Cu_{0.12}$

Bi$_2$Se$_3$ was grown. High purity (5N) metals of Cu, Bi and Se were taken in the stoichiometric ratio in a point bottomed alumina crucible (6 cm long and 16 mm diameter). The crucible was then sealed in a quartz tube under vacuum (10$^{-6}$ torr) and placed inside the furnace where it was heated at 850°C above the melting point of this compound. The sample was held at this temperature for about 24 hours before it was cooled down at a rate of about 1C/hr down to 650°C at which point the ampoule was quenched in liquid nitrogen. Large size single crystal of about 1.5 cm long was obtained and the crystal was easily cleavable with the cleavage place corresponding to c-plane of the crystal. Well defined Laue diffraction points confirmed the good quality of the single crystal.

A commercial SQUID magnetometer (MPMS 5, Quantum Design, USA) was used to measure the temperature-dependence of the magnetic susceptibility χ in a field of 50 Oe in the temperature range from 1.8 to 300 K to confirm the superconductivity exhibited in this compound. In addition to the magnetic susceptibility measurement, electrical resistivity was also measured in the four-probe geometry in a home made set up.

Since the crystal is brittle in nature so for TEM sample preparation a piece of crystal was placed in 1.2 mmX1.8 mm slot of Titanium 3 slots grid (Technoorg Linda, Hungary) and fixed with G1 epoxy (Gatan Inc,USA). Grinding and dimpling were done in the next steps to thin down the specimen up to a residual thickness of 10 to 15 μm. Finally double-sided Ar$^+$ beam milling was performed at small angles (< 6°) and at low energies (acceleration voltage: 2.5 kV; beam current < 8 μA) to avoid substantial heating of the TEM foils and consequently the introduction of artifacts.

The experiment was performed using a FEI-TITAN microscope equipped with FEG source, GIF 'Tridem' energy filter at an operating voltage of 300 kV. All experimental data were collected when the specimen was oriented at [-1-1 2 0] zone axis. For exit face wave retrieval a through focal series containing 8 images with a starting defocus of 45 nm and defocus step of 5 nm in underfocus direction were acquired in energy filtered TEM mode with an energy window of 10 eV around the zero loss peak to record elastic scattering related information. Experimental conditions were maintained to keep the complete imaged regions within parallel illumination. Images were acquired on a 2k X 2K CCD camera using Digital Micrograph software (Gatan Inc., USA). Gerchberg-Saxton method [8] in MactemPasX software (version 2.3.19) was used for exit face wave reconstruction.

To check whether the as grown single crystal $Cu_xBi_2Se_3$ (x = 0.12) exhibits superconductivity or not, electrical resistivity and magnetic susceptibility were measured. The main panel of Fig.1 shows electrical resistivity of $Cu_{0.12}Bi_2Se_3$ for the current perpendicular to the *c*-axis. A sharp drop in electrical resistivity at 3.1 K as shown in the top inset indicates the onset of superconductivity. Similarly, a diamagnetic transition at 3.1 K as shown in the lower inset of Fig.1 confirms the onset of superconductivity. The electrical resistivity and the magnetic susceptibility, data are almost similar to the previously published data by Hor et al[7].

To confirm the presence of intercalation the material was investigated further using TEM. Fig. 2a represents the selected area diffraction pattern. One image from the defocus series and the reconstructed phase of the exit face wave are presented in Figs. 2b and 2c respectively. A single TEM image does not contain any phase information which is material specific. This is only the intensity distribution in 2-D which is influenced by different parameters like image defocus, lens aberrations, specimen thickness etc. To retrieve the true material specific information the exit face wave series which is free from lens effects was reconstructed using through focal image. Phase change of the electron wave after passing through the sample can be expressed as $\Phi = V(r) \sigma t$, where $V(r)$ is the electrostatic potential seen by the incoming electron, $\sigma$ is the relativistic electron interaction constant and t is the specimen thickness along the electron beam. Using two images; one with an energy window of 10 eV around zero-loss peak and another without energy window (with the whole spectrum) specimen thickness along the electron beam was determined using equation (22) of a previous article[9]. For the present study a region of uniform thickness was chosen for exit face wave reconstruction so that the phase change actually encodes the change in potential. One of the line profiles (from bottom to top) which was taken from the region marked with white boundary in Fig. 2c is presented in Fig. 2d. This profile contains three groups of atomic columns according to the potential classified as 1, 2 and 3. Ratio of average area under the peaks of 1 and 2 matches with the ratio of electron scattering potentials of Bi and Se calculated from literature[10] which reveals that atomic columns 1 and 2 contain Bi and Se respectively. Schematic of $Bi_2Se_3$ structure (as proposed by[11]) projected along [-1-1 2 0] direction is presented in Fig. 3a. Comparing Fig. 2d and the atomic columns within the red boundary of Fig. 3a it can be stated that there are two extra atomic sites (marked as 3) present in the studied material when viewed along [-1-1 2 0]. Along the directions marked as A and B the ratio of nearest neighbor Bi to Bi distance in Fig 2c (~3.3) is elongated than Fig 3a (~2.6) though the angle between them remains the same which also indirectly supports the presence of intercalation sites along A. Similar to the previous report [7] assuming the presence of Cu atoms in intercalation sites and using both the ratio of scattering factor ( Bi or Se and Cu) and area under the peaks (1 or2 and 3) the calculated stoichiometry of the intercalated unit cell turns out to be $CuBi_2Se_3$.

Schematic of $Cu_xBi_2Se_3$ structure (as proposed by [7]) projected along [-1-1 2 0] direction is presented in Fig. 3b. Comparing Figs. 2c and 3b it can be stated that the position of intercalation sites are different for these two structures. The experimental data exhibits two intercalation sites in direction marked as A in Fig. 2c but the previously proposed structure [7] shows one intercalation site in the direction marked as C in Fig 3b. Therefore, the structure of intercalated unit cell needs to be studied further.

To conclude, the present study provides direct evidence of intercalation sites in a doped topological insulator turned superconductor material ($Cu_xBi_2Se_3$). Results show two intercalation sites in this material when projected along the [-1-1 2 0] direction. Assuming Cu intercalation calculated stoichiometry of the intercalated unit cell was found to be $CuBi_2Se_3$. It was also seen that the earlier proposed structure of this material [7] does not match with the present result.

## Acknowledgement

Heartfelt thanks are expressed to Mr. Jayesh B. Parmar for preparing TEM samples.

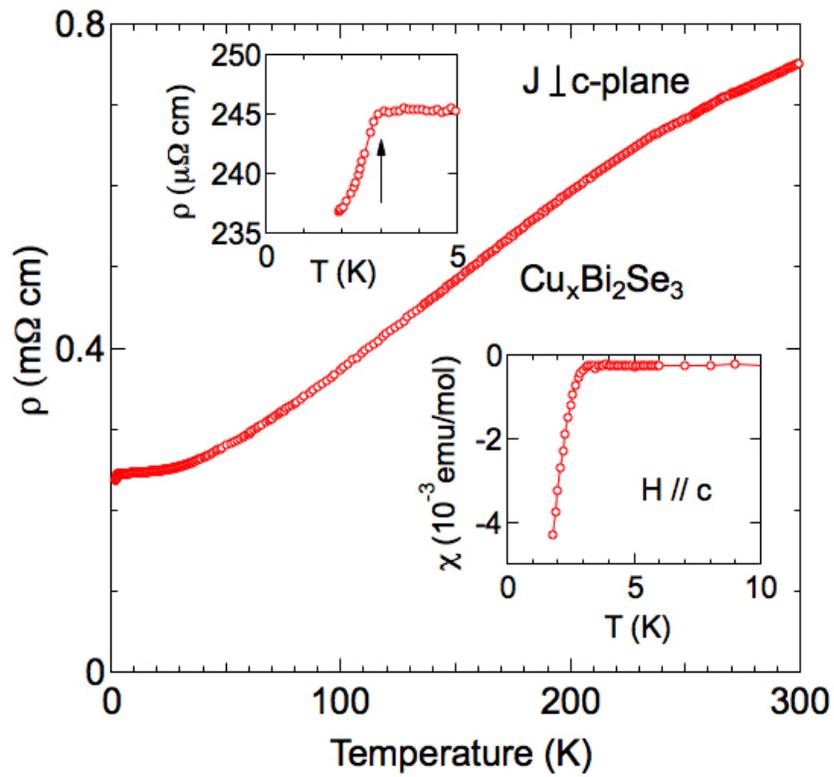

**Fig. 1.** Temperature dependence of electrical resistivity for current perpendicular to the *c*-plane. The top inset shows the low temperature resistivity data and the bottom inset shows the clear diamagnetic signal, confirming the superconducting transition.

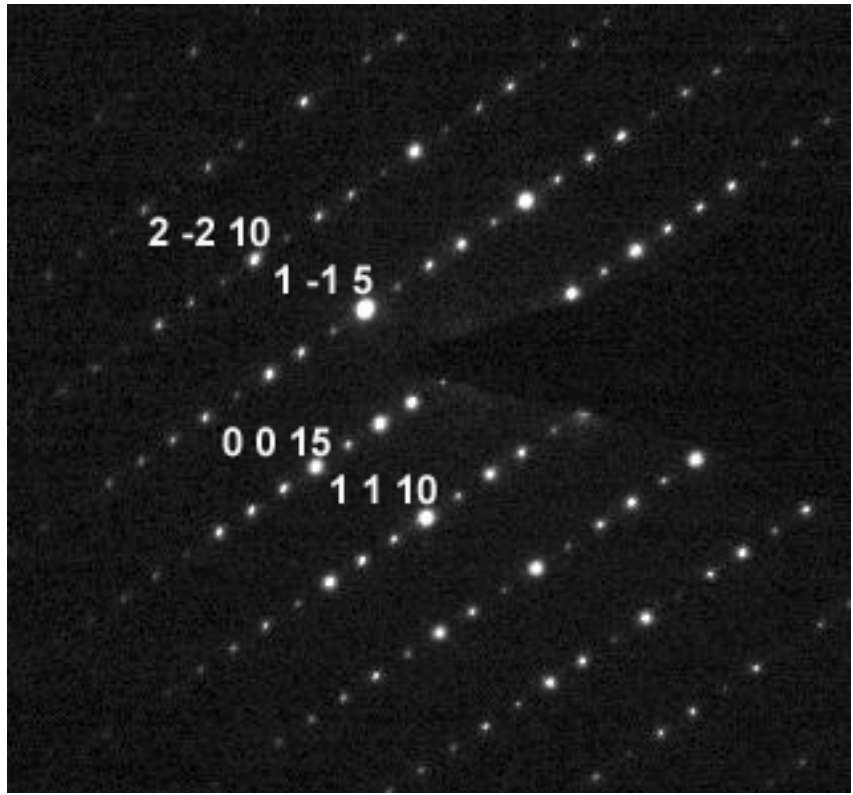

**Fig. 2.** a. Selected area diffraction pattern of CuxBi2Se3 along [-1-1 2 0] direction. Diffraction spots are marked using three index system.

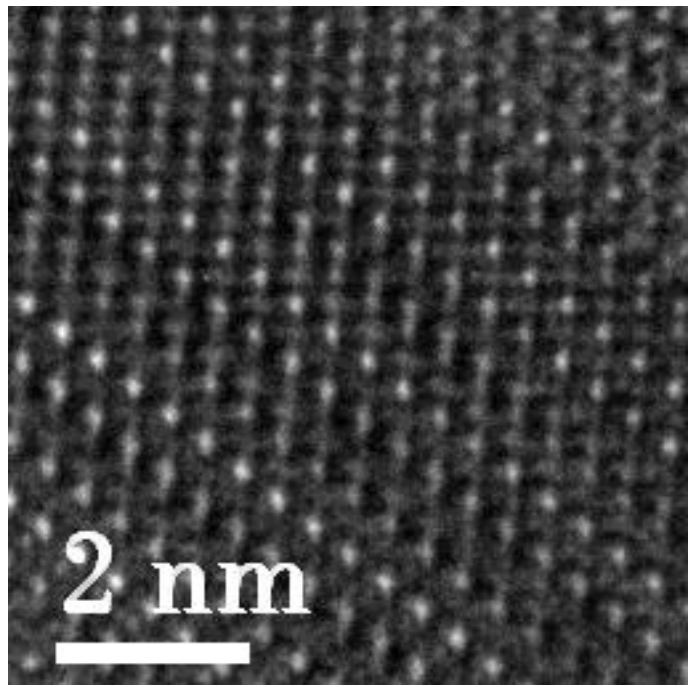

**Fig. 2** b. A high resolution lattice image of the sample at same viewing direction.

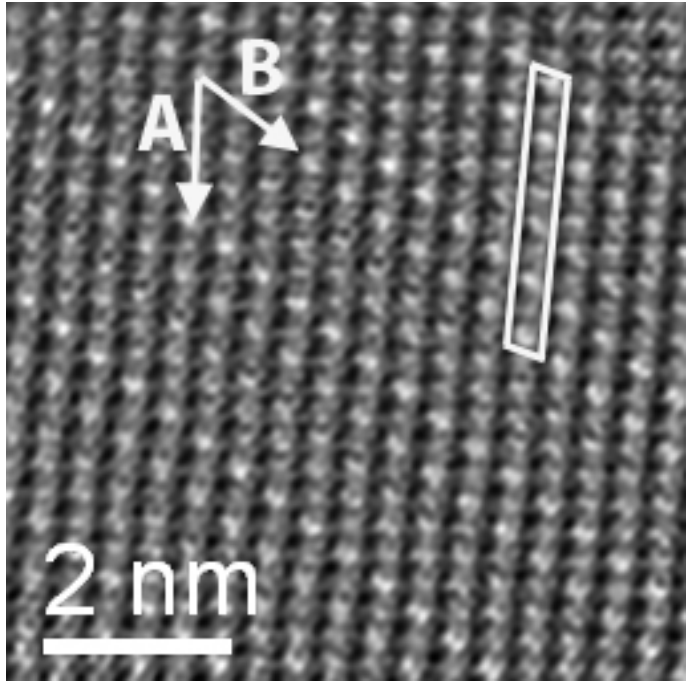

**Fig. 2** c. Reconstructed phase of the region shown in Fig. 2b.

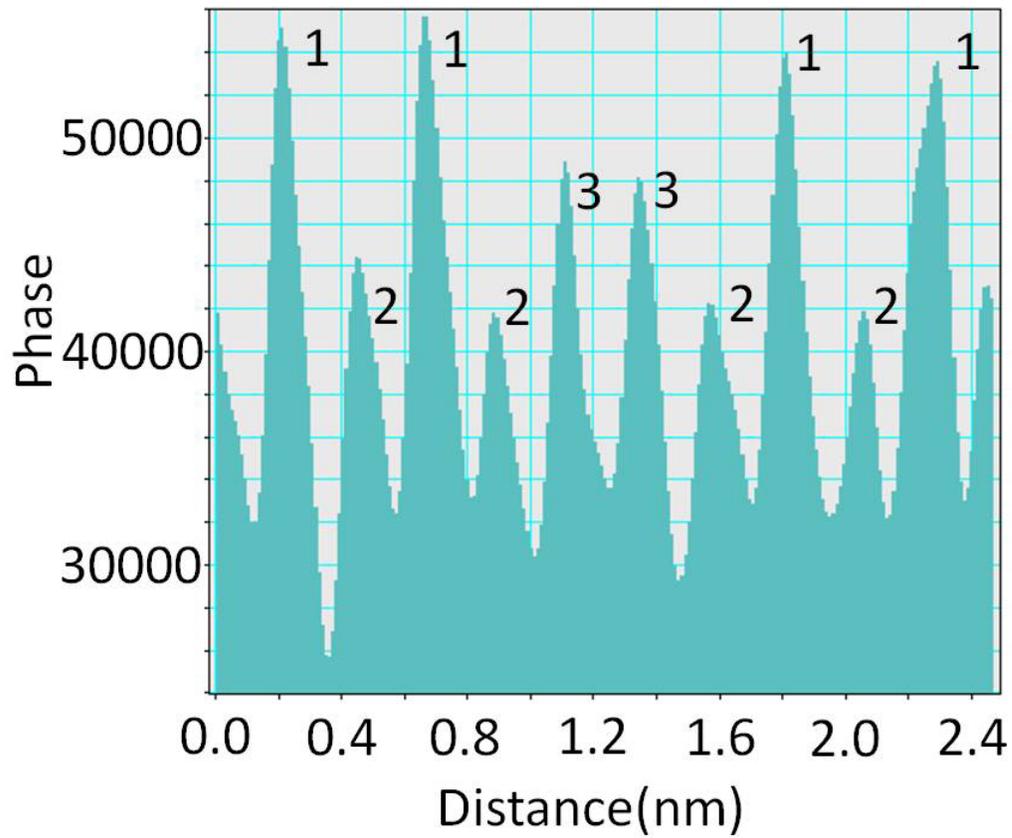

**Fig. 2** d. Line profile (from bottom to top) taken from the region marked with white boundary in Fig. 2c.

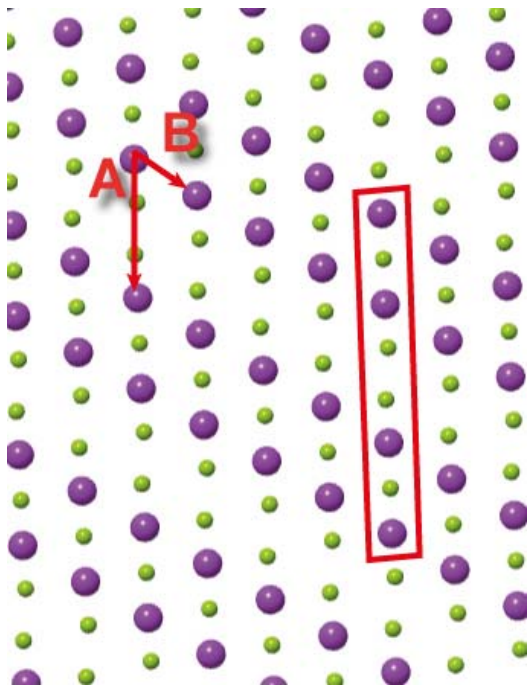

**Fig. 3** a. Schematic of undoped $Bi_2Se_3$ structure (as proposed by [7] ) projected along [-1-1 2 0] direction.

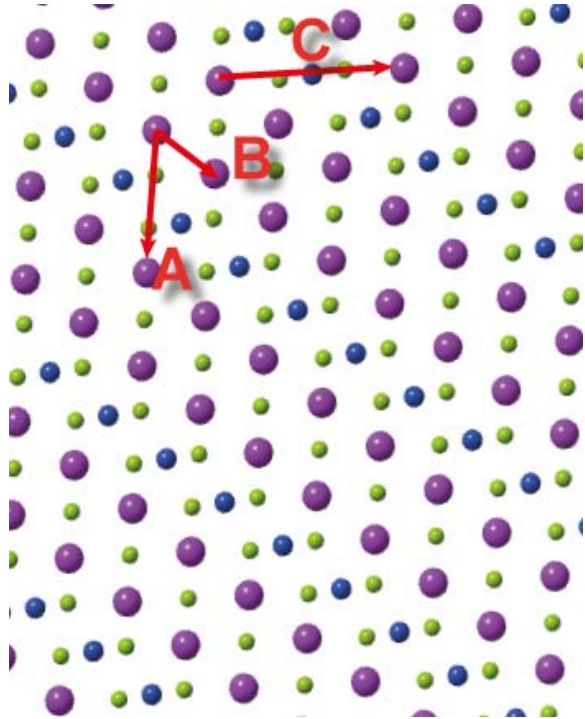

**Fig. 3** b. Schematic of $Cu_xBi_2Se_3$ structure (as proposed by [7] ) projected along [-1-1 2 0] direction. Bi, Se and Cu atoms are represented as purple, green and blue colored circles respectively.